\documentstyle[titlepage,12pt]{article}

\begin{document}
\title{Scattering of string-waves on black hole background}
\vspace{5 in}
\author{A.L.Larsen\thanks{E-mail: allarsen@nbivax.nbi.dk}\\Nordita,
Blegdamsvej 17, DK-2100 Copenhagen \O, Denmark}
\maketitle

\begin{abstract}
We consider the propagation of perturbations along an infinitely long
stationary open string in the background of a Schwarzschild black hole.
The equations of motion for the perturbations in the 2 transverse physical
directions are solved to second order in a weak field expansion. We then
set up a scattering formalism where an ingoing wave is partly transmitted
and partly reflected due to the interaction with the gravitational field
of the black hole. We finally calculate the reflection coefficients to third
order in our weak field expansion.
\end{abstract}
\newpage
\section{Introduction and conclusions}
In a previous publication [1] we developed a covariant formalism for physical
perturbations propagating along a string in an arbitrary curved background.
As applications we considered stationary strings in quasi-Newtonian, Rindler
and de Sitter spacetimes, and we presented the wave-equations determining the
evolution of the perturbations when propagating along a string in the
background of different kinds of black holes. In the latter cases, however,
solving these wave-equations explicitly was not attempted, and therefore the
physical quantities (scattering amplitudes,...) describing the propagation of
the waves were not extracted from the general formalism.

In the present paper we return to these wave-equations in the special case of
perturbations propagating along an infinitely long stationary string in the
equatorial plane of a Schwarzschild black hole. We make an expansion in the
dimensionless parameter $\lambda\equiv r_g/b$ [2], where $r_g$ is the
gravitational radius of the black hole and $b$ is the "impact parameter"
measuring the minimal distance between the black hole and the stationary
string. We then solve the wave-equations up to second order in $\lambda$,
thus obtaining explicit expressions for the physical perturbations
$\delta x^{(0-2)}_\parallel(\sigma_c,\tau)$ and
$\delta x^{(0-2)}_\perp(\sigma_c,\tau)$ parallel and
perpendicular to the equatorial plane of the black hole, respectively
($\tau$ and $\sigma_c$ are the 2 world-sheet coordinates).
Far away from the black hole the geometry of spacetime is flat and the string
perturbations are simple plane waves. We can then set up a scattering formalism
with a localized regime near the black hole where the scattering takes place
and 2 "free" asymptotic regions. We derive the transformation matrix relating
the "out"-amplitudes to the "in"-amplitudes and we calculate the reflection
coefficients for the 2 physical polarizations up to third order in $\lambda$.
The reflection coefficient is largest for perturbations in the equatorial
plane, but in both cases it is only appreciably different from zero when
the wavelength at infinity $\lambda_\infty$ is comparable to the
impact parameter $b$.
Although the physical situation is considerably different it turns out that
our analysis is somewhat similar to the analysis of the quantum scattering of
fundamental strings by black holes carried out by Vega and Sanchez [2-3].

The paper is organized as follows: In section 2 we give a very short review
of the general formalism developed in Ref.1., and in section 3 we use this
general formalism to derive the wave-equations for perturbations
propagating along a stationary string in the equatorial plane of a
Schwarzschild black hole. In section 4 we introduce the weak field
approximation [2] where the perturbations and the potentials are expanded in
powers of the small parameter $\lambda$, and in sections 5-6 we solve the
wave-equations up to second order. Finally in section 7 we consider a
scattering proces and the reflection coefficients are calculated up to third
order in $\lambda$.

We use sign-conventions of Misner-Thorn-Wheeler [4] and units where $G=1$,
$c=1$ and the string tension $(2\pi\alpha')^{-1}=1$.
\newpage
\section{General approach}
In this section we give a short review of the derivation of the scalar
wave-equations determining the propagation of perturbations in the transverse
directions of a stationary string in a static background. Starting with the
Polyakov action [5]:
\begin{equation}
{\cal S}=\int d\tau d\sigma\sqrt{-h}h^{AB}G_{AB},
\end{equation}
and the corresponding equations of motion ($\Box\equiv(-h)^{-1/2}\partial_A
(\sqrt{-h}h^{AB}\partial_B)$):
\begin{equation}
G_{AB}=\frac{1}{2}h_{AB}G^C\hspace*{.5mm}_C,
\end{equation}
\begin{equation}
\Box x^\mu+h^{AB}\Gamma^\mu_{\rho\sigma}x^\rho_{,A}x^\sigma_{,B}=0,
\end{equation}
we consider variations/perturbations:
\begin{equation}
\delta h_{AB},\hspace*{5mm}\delta x^\mu=\delta x^R n^\mu_R
\end{equation}
around an exact solution. $n^\mu_R\hspace*{.5mm}(R=2,3)$ are 2 vectors normal
to the surface of the world-sheet of the exact solution, so we only consider
perturbations $\delta x^\mu$ in the physical directions.

Variation of Eq. (2.2) gives $\delta h_{AB}$ in terms of $\delta x^R$, which,
when used in the variation of Eq. (2.3), leads to a complicated matrix
wave-equation for the 2 physical perturbations $\delta x^R$. The details can be
found in Ref.1., here we just recall the result:
\begin{eqnarray}
\Box\delta x_R\hspace*{-2mm}&+&\hspace*{-2mm}2\mu_{RS}\hspace*{1mm}^A
(\delta x^S)_{,A}+(\nabla_A\mu_{RS}\hspace*{1mm}^A)\delta x^S-
\mu_{RT}\hspace*{1mm}^A\mu_S\hspace*{.5mm}^T\hspace*{.5mm}_A\delta x^S
\nonumber\\
\hspace*{-2mm}&+&\hspace*{-2mm}\frac{2}{G^C\hspace*{.5mm}_C}\Omega_R
\hspace*{.5mm}^{AB}\Omega_{S,AB}\delta x^S-h^{AB}x^\mu_{,A}x^\nu_{,B}
R_{\mu\rho\sigma\nu}n^\rho_R n^\sigma_S \delta x^S=0,
\end{eqnarray}
where $\nabla_A$ is the world-sheet covariant derivative, $R_{\mu\rho
\sigma\nu}$ is the spacetime Riemann-tensor and $\Omega_{R,AB}$ and
$\mu_{RS,A}$ are the second fundamental form and the normal fundamental
form, respectively [6]:
\begin{equation}
\Omega_{R,AB}=g_{\mu\nu}n^\mu_R x^\rho_{,A}\nabla_\rho x^\nu_{,B},
\end{equation}
\begin{equation}
\mu_{RS,A}=g_{\mu\nu}n^\mu_R x^\rho_{,A}\nabla_\rho n^\nu_S,
\end{equation}
where $\nabla_\rho$ is the spacetime covariant derivative.

Equation (2.5) holds for an arbitrary string configuration in an arbitrary
curved spacetime. In the case of a stationary string in a static background
some simplifications arise. The metric of a static spacetime is conveniently
written:
\begin{equation}
g_{\mu\nu}=\left( \begin{array}{cc}-F&0\\0&H_{ij}/F\end{array}\right),
\end{equation}
where $\partial_t F=0$, $\partial_t H_{ij}=0$; $i,j=1,2,3$. A stationary string
can be parametrized by:
\begin{equation}
t=x^0=\tau,\hspace*{5mm}x^i=x^i (\sigma); i=1,2,3.
\end{equation}
Using Eqs. (2.8)-(2.9) it can be shown that Eq. (2.5) reduces to
(in the conformal gauge):
\begin{equation}
(\partial^2_{\sigma_c}-\partial^2_\tau)\delta x_R=U_{RS}\delta x^S,
\end{equation}
where the matrix potential $U_{RS}$ is given by:
\begin{equation}
U_{RS}=V\delta_{RS}+F^{-1}V_{RS};
\end{equation}
\begin{equation}
V=\frac{3}{4F^2}\left(\frac{dF}{d\sigma_c}\right)^2-\frac{1}{2F}
\frac{d^2F}{d\sigma^2_c},
\end{equation}
\begin{equation}
V_{RS}=\frac{dx^i}{d\sigma_c}\frac{dx^j}{d\sigma_c}\tilde{R}_{iklj}n^k_R
n^l_S.
\end{equation}
Here $\tilde{R}_{iklj}$ is the Riemann tensor for the metric $H_{ij}$ and
$n^k_R$ represents the space components of the normalvectors introduced in Eq.
(2.4). For the details we refer the reader to Ref.1.
\section{Schwarzschild black hole}
\setcounter{equation}{0}
The results of Eqs. (2.10)-(2.13) in the special case of perturbations
propagating along a stationary string in the equatorial plane of a
Schwarzschild black hole were stated in Ref.1. without any details of the
derivation. In this section we show the detailed calculations and in the
following sections we then come to the solutions of the wave-equations (2.10).

The Schwarzschild black hole is given by the line element:
\begin{equation}
ds^2=-(1-2m/r)dt^2+(1-2m/r)^{-1}dr^2+r^2(d\theta^2+\sin^2\theta d\phi^2),
\end{equation}
where $m$ is the mass of the black hole. Comparing with Eq. (2.8) we find:
\begin{equation}
F=1-2m/r,
\end{equation}
\begin{equation}
H_{ij}=diag(1,\Delta,\Delta\sin^2\theta);\hspace*{4mm}\Delta\equiv r^2-2mr.
\end{equation}
The Christoffel symbols $\tilde{\Gamma}^i_{jk}$ for the metric $H_{ij}$ are
given by:
\begin{eqnarray}
\tilde{\Gamma}^r_{\phi\phi}\hspace*{-2mm}&=&\hspace*{-2mm}\sin^2\theta
\tilde{\Gamma}^r_{\theta\theta}=-(r-m)\sin^2\theta,\nonumber\\
\tilde{\Gamma}^\theta_{r\theta}\hspace*{-2mm}&=&\hspace*{-2mm}
\tilde{\Gamma}^\phi_{r\phi}=\frac{r-m}{\Delta},\nonumber\\
\tilde{\Gamma}^\theta_{\phi\phi}\hspace*{-2mm}&=&\hspace*{-2mm}-\cos\theta
\sin\theta,\hspace*{5mm}\tilde{\Gamma}^\phi_{\phi\theta}=\cot\theta,
\end{eqnarray}
and the non-vanishing components of the Riemann tensor $\tilde{R}_{iklj}$ are:
\begin{eqnarray}
&\tilde{R}_{r\phi r\phi}=\sin^2\theta\tilde{R}_{r\theta r\theta}=\sin^2\theta
\frac{m^2}{\Delta},&\nonumber\\
&\tilde{R}_{\theta\phi\theta\phi}=-m^2\sin^2\theta.&
\end{eqnarray}
Let us now consider the stationary string in the background of the black hole
given by the line element (3.1). Taking the string in the equatorial plane and
still working in the
conformal gauge, the configuration is determined by [7]:
\begin{equation}
t=\tau,\hspace*{5mm}\theta=\pi/2,
\end{equation}
\begin{equation}
\frac{d\phi}{d\sigma_c}=\frac{b}{r^2},
\end{equation}
\begin{equation}
\frac{dr}{d\sigma_c}=\pm\sqrt{(1-\frac{2m}{r})^2-\frac{b^2}{r^2}
(1-\frac{2m}{r})},
\end{equation}
where $b$ is an integration constant. The 2 normalvectors introduced in Eq.
(2.4) are then chosen in the following way:
\begin{equation}
n^0_2\equiv n^0_\perp=0,\hspace*{5mm}n^0_3\equiv n^0_\parallel=0,
\end{equation}
\begin{equation}
n^k_2\equiv n^k_\perp=\frac{1}{r}(0,1,0),
\end{equation}
\begin{equation}
n^k_3\equiv n^k_\parallel=\frac{1}{r}(-b,0,\frac{1}{1-2m/r}\frac{dr}
{d\sigma_c}).
\end{equation}
{}From Eqs. (2.12)-(2.13), (3.2)-(3.5) and (3.7)-(3.11) we find:
\begin{eqnarray}
&V=\frac{m}{\Delta r^4}[\Delta(2r-3m)+(4m-3r)b^2],&\nonumber\\
&V_{\perp\parallel}=V_{\parallel\perp}=0,&\nonumber\\
&V_{\perp\perp}=\frac{m^2}{r^6}(2b^2-\Delta),\hspace*{5mm}
V_{\parallel\hspace*{.5mm}\parallel}=-\frac{m^2}{r^6}\Delta.&
\end{eqnarray}
Eq. (2.10) then leads to the 2 decoupled equations [1]:
\begin{equation}
(\partial^2_{\sigma_c}-\partial^2_\tau)\delta x_\perp=\frac{m}{r^5}
(2\Delta-3b^2)\delta x_\perp\equiv U_{\perp\perp}\delta x_\perp,
\end{equation}
\begin{equation}
(\partial^2_{\sigma_c}-\partial^2_\tau)\delta x_\parallel=[\frac{m}{r^5}
(2\Delta-3b^2)-\frac{2m^2b^2}{\Delta r^4}]\delta x_\parallel\equiv
U_{\parallel\hspace*{.5mm}\parallel}\delta x_\parallel,
\end{equation}
determining the evolution of the 2 physical polarizations of perturbations
(perpendicular and parallel to the equatorial plane). The way to proceed now
is to solve Eq. (3.8) for $r(\sigma_c)$ and then to write $U_{\perp\perp}$ and
$U_{\parallel\hspace*{.5mm}\parallel}$ as explicit functions of $\sigma_c$.
Finally we then have to solve Eqs. (3.13)-(3.14) for $\delta x_\perp
(\sigma_c,\tau)$ and $\delta x_\parallel (\sigma_c,\tau)$.
\section{Weak field expansion}
\setcounter{equation}{0}
Actually solving Eqs. (3.13)-(3.14) turns out to be a very difficult task.
Equation (3.8) can of course be solved for $r(\sigma_c)$ in terms of elliptic
(or hyperelliptic) functions, and the potentials $U_{\perp\perp}$ and
$U_{\parallel\hspace*{.5mm}\parallel}$ are then some complicated rational
expressions in these elliptic functions. The shape of the potentials are
sketched in Fig.1. Far away from the black hole the potentials are obviously
flat but there is a well surrounded by potential barriers where the string is
closest to the black hole. The potential for perturbations in the equatorial
plane is a little deeper than for perturbations perpendicular to the equatorial
plane, but otherwise they have the same shape.

We have not been able to find the general analytic solution to Eqs.
(3.13)-(3.14). They can of course easily be solved nurmerically, but we will
instead consider approximate analytic solutions based on a weak field
expansion [2]. First note that the integration constant $b$ introduced in Eqs.
(3.7)-(3.8) is related to the minimal distance between the infinitely long
open string and the black hole by [7]:
\begin{equation}
r_{min}=m+\sqrt{m^2+b^2}.
\end{equation}
$b$ thus plays the role of an "impact parameter". For $b=0$ the string touches
the horizon of the black hole while for $b\neq 0$ it is always strictly
outside. We then define our dimensionless expansion parameter $\lambda$ by:
\begin{equation}
\lambda\equiv\frac{r_{g}}{b},
\end{equation}
where $r_g=2m$ is the gravitational radius of the black hole. If $\lambda$
is small the string is far away from the black hole, corresponding to a weak
field approximation [2]. In the following we will consider $b$ to be a finite
positive constant so that the zeroth order in
$\lambda$ is obtained by $r_g=0$ (and
not $b=\infty$!).

We now expand Eqs. (3.13)-(3.14) in powers of $\lambda$. First consider the
coordinates $(r,\phi)$ of the stationary string:
\begin{equation}
r=r_{(0)}+r_{(1)}+r_{(2)}+...,
\end{equation}
\begin{equation}
\phi=\phi_{(0)}+\phi_{(1)}+\phi_{(2)}+...
\end{equation}
{}From Eqs. (3.7)-(3.8) we find to zeroth order in $\lambda$:
\begin{equation}
\frac{d\phi_{(0)}}{d\sigma_c}=\frac{b}{r^2_{(0)}},
\end{equation}
\begin{equation}
\frac{dr_{(0)}}{d\sigma_c}=\pm\sqrt{1-\frac{b^2}{r^2_{(0)}}}.
\end{equation}
These 2 equations are easily solved by:
\begin{equation}
r_{(0)}=\sqrt{\sigma^2_c+b^2},
\end{equation}
\begin{equation}
\phi_{(0)}=\arctan\frac{\sigma_c}{b},
\end{equation}
i.e. it is just the straight string. This is of course not surprising since
at the zeroth order there is no black hole at all, and the only stationary
string in flat spacetime is the straight one. At first order in $\lambda$
we find:
\begin{equation}
\frac{dr_{(1)}}{d\sigma_c}=\pm\left( \frac{b^2-2r^2_{(0)}}{2r^2_{(0)}
\sqrt{r^2_{(0)}-b^2}}r_g+
\frac{b^2}{r^2_{(0)}\sqrt{r^2_{(0)}-b^2}}r_{(1)}\right),
\end{equation}
\begin{equation}
\frac{d\phi_{(1)}}{d\sigma_c}=-\frac{2b}{r^3_{(0)}}r_{(1)}.
\end{equation}
Using Eq. (4.7) these 2 equations are solved by:
\begin{equation}
r_{(1)}=r_g(\frac{1}{2}-\frac{\sigma_c}{\sqrt{\sigma^2_c+b^2}}\sinh^{-1}
\frac{\sigma_c}{b}),
\end{equation}
\begin{equation}
\phi_{(1)}=-\frac{br_g}{\sigma^2_c+b^2}\sinh^{-1}\frac{\sigma_c}{b},
\end{equation}
where the integration constants have been chosen such that:
\begin{equation}
\phi(0)=0,\hspace*{5mm}\frac{dr}{d\sigma_c}(0)=0.
\end{equation}
We can then continue to solve Eqs. (3.7)-(3.8) order by order, but for our
purposes we will only need $r$ and $\phi$ up to first order in $\lambda$.

Now consider the 2 potentials $U_{\perp\perp}$ and $U_{\parallel\hspace*{.5mm}
\parallel}$. Obviously the potentials have no zeroth order contributions (flat
spacetime), are identical at first order but in general different at second
and higher orders. We find:
\begin{equation}
\frac{m}{r^5}(2\Delta-3b^2)=\frac{2r^2_{(0)}-3b^2}{2r^5_{(0)}}r_g-
\frac{r^2_g}{r^4_{(0)}}+3\frac{5b^2-2r^2_{(0)}}{2r^6_{(0)}}r_g r_{(1)}
+{\cal O}(\lambda^3),
\end{equation}
\begin{equation}
-\frac{2m^2b^2}{\Delta r^4}=-\frac{b^2 r^2_g}{2r^6_{(0)}}+{\cal O}(\lambda^3).
\end{equation}
Using Eqs. (4.7) and (4.11) these 2 expressions give the
potentials $U_{\perp\perp}$ and $U_{\parallel\hspace*{.5mm}\parallel}$ as
explicit functions of $\sigma_c$ up to second order in $\lambda$.

Finally we also expand the perturbations propagating along the stationary
string:
\begin{equation}
\delta x_\perp=\delta x^{(0)}_\perp+\delta x^{(1)}_\perp+
\delta x^{(2)}_\perp+...,
\end{equation}
\begin{equation}
\delta x_\parallel=\delta x^{(0)}_\parallel+\delta x^{(1)}_\parallel+
\delta x^{(2)}_\parallel+...
\end{equation}
Collecting everything we can then write down the wave-equations (3.13)-(3.14)
up to the second order in $\lambda$.
\section{Zeroth and first order equations and their solutions}
\setcounter{equation}{0}
We now come to the solutions of the wave-equations (2.10) up to first order
in the expansion described in the previous section. To zeroth order we have:
\begin{equation}
(\partial^2_{\sigma_c}-\partial^2_\tau)\delta x^{(0)}_\perp=
(\partial^2_{\sigma_c}-\partial^2_\tau)\delta x^{(0)}_\parallel=0,
\end{equation}
i.e. it is just the ordinary flat spacetime wave-equations (as it should be!).
The solutions are generally written as integrals over the continuous frequency
plane waves:
\begin{equation}
\delta x^{(0)}_\perp(\sigma_c,\tau)=\int d\omega\left( a^\perp_\omega
e^{-i\omega(\tau-\sigma_c)}+
b^\perp_\omega e^{-i\omega(\tau+\sigma_c)}\right),
\end{equation}
\begin{equation}
\delta x^{(0)}_\parallel(\sigma_c,\tau)=\int d\omega\left( a^\parallel_\omega
e^{-i\omega(\tau-
\sigma_c)}+b^\parallel_\omega e^{-i\omega(\tau+\sigma_c)}\right),
\end{equation}
where:
\begin{eqnarray}
&(a^\perp_\omega)^\ast=a^\perp_{-\omega},\hspace*{5mm}(a^\parallel_\omega)
^\ast=a^\parallel_{-\omega},&\nonumber\\
&(b^\perp_\omega)^\ast=b^\perp_{-\omega},\hspace*{5mm}(b^\parallel_\omega)
^\ast=b^\parallel_{-\omega}.&\nonumber
\end{eqnarray}
The first order wave-equations are:
\begin{equation}
(\partial^2_{\sigma_c}-\partial^2_\tau)\delta x^{(1)}_\perp=U^{(1)}_
{\perp\perp}\delta x^{(0)}_\perp,
\end{equation}
\begin{equation}
(\partial^2_{\sigma_c}-\partial^2_\tau)\delta x^{(1)}_\parallel=
U^{(1)}_{\parallel\hspace*{.5mm}\parallel}\delta x^{(0)}_\parallel,
\end{equation}
where we used the fact that the potentials have no zeroth order terms. From
Eqs. (3.13)-(3.14) and (4.14)-(4.15) we find:
\begin{equation}
U^{(1)}_{\perp\perp}=U^{(1)}_{\parallel\hspace*{.5mm}\parallel}=
\frac{2r^2_{(0)}-3b^2}{2r^5_{(0)}}r_g=m\frac{2\sigma^2_c-b^2}
{(\sigma^2_c+b^2)^{5/2}}.
\end{equation}
To proceed it is convenient to Fourier-expand the perturbations
$\delta x^{(1)}_\perp$ and $\delta x^{(1)}_\parallel$:
\begin{equation}
\delta x^{(1)}_\perp(\sigma_c,\tau)=\int d\omega D^\perp_\omega (\sigma_c)
e^{-i\omega\tau},
\end{equation}
\begin{equation}
\delta x^{(1)}_\parallel(\sigma_c,\tau)=\int d\omega D^\parallel_\omega
(\sigma_c)e^{-i\omega\tau}.
\end{equation}
Now equations (5.4)-(5.5) lead to:
\begin{equation}
(\frac{d^2}{d\sigma^2_c}+\omega^2)D^R_\omega (\sigma_c)=m\frac{2\sigma^2_c-
b^2}{(\sigma^2_c+b^2)^{5/2}}\left(a^R_\omega e^{i\omega\sigma_c}+
b^R_\omega e^{-i\omega\sigma_c}\right),
\end{equation}
where $R$ is either $"\perp"$ or $"\parallel"$. This equation is solved by:
\begin{equation}
D^R_\omega (\sigma_c)=e^{i\omega\sigma_c}\left(a^R_\omega A_\omega
(\sigma_c)+b^R_\omega B^\ast_\omega(\sigma_c)\right)+e^{-i\omega\sigma_c}
\left( a^R_\omega B_\omega (\sigma_c)+b^R_\omega A^\ast_\omega
(\sigma_c)\right),
\end{equation}
where:
\begin{equation}
A_\omega (\sigma_c)=\frac{-m}{2i\omega}\frac{\sigma_c}{(\sigma^2_c+b^2)^{3/2}},
\end{equation}
\begin{equation}
B_\omega(\sigma_c)=\frac{-m}{2i\omega}\int^{\sigma_c}_{-\infty}
\frac{2x^2-b^2}{(x^2+b^2)^{5/2}}e^{2i\omega x}dx,
\end{equation}
and we have imposed the boundary conditions:
\begin{equation}
D^\perp_\omega(-\infty)=D^\parallel_\omega(-\infty)=0.
\end{equation}
The first order perturbations are finally given by:
\begin{eqnarray}
\delta x^{(1)}_R(\sigma_c,\tau)=\int d\omega[\hspace*{-4mm}&
e^{-i\omega(\tau-\sigma_c)}
(a^R_\omega A_\omega(\sigma_c)+
b^R_\omega B^\ast_\omega(\sigma_c))&\nonumber\\
\hspace*{-5mm}&+e^{-i\omega(\tau+\sigma_c)}(a^R_\omega B_\omega (\sigma_c)+
b^R_\omega A^\ast_\omega (\sigma_c))]&\hspace*{-2mm},
\end{eqnarray}
Note that $A_\omega(\sigma_c)$ and $B_\omega(\sigma_c)$ are
independent of the physical polarization, so that up to this order in $\lambda$
there is no difference in the way the perturbations propagate in the 2
physical directions.
\section{Second order equation and its solution}
\setcounter{equation}{0}
At second order in $\lambda$ things become quite complicated. It is however
important to calculate the second order corrections to the plane wave
perturbations given by Eqs. (5.2)-(5.3), since this is the order where the
perturbations can
really begin to reflect the geometry in the sense that this is the lowest
order where we can possibly find a difference between the perturbations in
the 2 physical directions. For waves on a string in the
equatorial plane of a black hole we obviously expect some differences for the
propagation of perturbations in these 2 directions. The second order
wave-equations are:
\begin{equation}
(\partial^2_{\sigma_c}-\partial^2_\tau)\delta x^{(2)}_\perp=U^{(2)}
_{\perp\perp}\delta x^{(0)}_\perp+U^{(1)}_{\perp\perp}\delta x^{(1)}_\perp,
\end{equation}
\begin{equation}
(\partial^2_{\sigma_c}-\partial^2_\tau)\delta x^{(2)}_\parallel=
U^{(2)}_{\parallel\hspace*{.5mm}\parallel}\delta x^{(0)}_\parallel+
U^{(1)}_{\parallel\hspace*{.5mm}\parallel}\delta x^{(1)}_\parallel.
\end{equation}
So we need explicit expressions for $U^{(2)}_{\perp\perp}$ and
$U^{(2)}_{\parallel\hspace*{.5mm}\parallel}$. From Eqs. (3.13)-(3.14) and
(4.14)-(4.15) we find:
\begin{eqnarray}
U^{(2)}_{\perp\perp}\hspace*{-2mm}&=&\hspace*{-2mm}3\frac{5b^2-2r^2_{(0)}}
{2r^6_{(0)}}r_g r_{(1)}-\frac{r^2_g}{r^4_{(0)}}\nonumber\\
\hspace*{-2mm}&=&\hspace*{-2mm}6m^2\frac{3b^2-2\sigma^2_c}
{(\sigma^2_c+b^2)^3}(\frac{1}{2}-
\frac{\sigma_c}{\sqrt{\sigma^2_c+b^2}}\sinh^{-1}\frac{\sigma_c}{b})-
\frac{4m^2}{(\sigma^2_c+b^2)^2},
\end{eqnarray}
as well as:
\begin{equation}
U^{(2)}_{\parallel\hspace*{.5mm}\parallel}=U^{(2)}_{\perp\perp}-\frac{b^2
r^2_g}{2r^6_{(0)}}=U^{(2)}_{\perp\perp}-\frac{2m^2b^2}{(\sigma^2_c+b^2)^3},
\end{equation}
where also Eq. (4.11) was used. To solve Eqs. (6.1)-(6.2) we Fourier-expand
the perturbations $\delta x^{(2)}_\perp$ and $\delta x^{(2)}_\parallel$:
\begin{equation}
\delta x^{(2)}_\perp(\sigma_c,\tau)=\int d\omega E^\perp_\omega(\sigma_c)
e^{-i\omega\tau},
\end{equation}
\begin{equation}
\delta x^{(2)}_\parallel(\sigma_c,\tau)=\int d\omega E^\parallel_\omega
(\sigma_c) e^{-i\omega\tau}.
\end{equation}
Using the results of section 5, Eqs. (6.1)-(6.2) reduce to
(no summation over $R$):
\begin{eqnarray}
(\frac{d^2}{d\sigma^2_c}+\omega^2)E^R_\omega(\sigma_c)\hspace*{-2mm}&=&
\hspace*{-2mm}(a^R_\omega f^R_\omega(\sigma_c)+
b^R_\omega(g^R_\omega(\sigma_c))^\ast)e^{i\omega\sigma_c}\nonumber\\
\hspace*{-2mm}&+&\hspace*{-2mm}(a^R_\omega g^R_\omega(\sigma_c)+
b^R_\omega(f^R_\omega(\sigma_c))^\ast)e^{-i\omega\sigma_c},
\end{eqnarray}
where $R$ is either $"\perp"$ or $"\parallel"$, and the functions $f^R_\omega$
and $g^R_\omega$ are given by:
\begin{equation}
f^R_\omega=U^{(2)}_{RR}+A_\omega U^{(1)}_{RR},
\end{equation}
\begin{equation}
g^R_\omega=B_\omega U^{(1)}_{RR}.
\end{equation}
The solution to equation (6.7) is:
\begin{equation}
E^R_\omega(\sigma_c)=e^{i\omega\sigma_c}\left(a^R_\omega X^R_\omega(\sigma_c)+
b^R_\omega(Y^R_\omega(\sigma_c))^\ast\right)+e^{-i\omega\sigma_c}
\left(a^R_\omega Y^R_\omega(\sigma_c)+b^R_\omega(X^R_\omega(\sigma_c))^\ast
\right)
\end{equation}
where:
\begin{equation}
X^R_\omega(\sigma_c)=\frac{1}{2i\omega}\int_{-\infty}^{\sigma_c}(f^R_\omega(x)
+e^{-2i\omega x}g^R_\omega (x))dx,
\end{equation}
\begin{equation}
Y^R_\omega(\sigma_c)=\frac{-1}{2i\omega}\int_{-\infty}^{\sigma_c}(g^R_\omega(x)+
e^{2i\omega x}f^R_\omega(x))dx,
\end{equation}
and we have imposed the boundary conditions:
\begin{equation}
E^\perp_\omega(-\infty)=E^\parallel_\omega(-\infty)=0.
\end{equation}
The second order perturbations are finally given by:
\begin{eqnarray}
\delta x^{(2)}_R (\sigma_c,\tau)=\int d\omega[\hspace*{-4mm}
&e^{-i\omega(\tau-\sigma_c)}(
a^R_\omega X^R_\omega (\sigma_c)+b^R_\omega(Y^R_\omega(\sigma_c))^\ast)&
\nonumber\\
\hspace*{-5mm}&+e^{-i\omega(\tau+\sigma_c)}(a^R_\omega
Y^R_\omega(\sigma_c)+b^R_\omega
(X^R_\omega(\sigma_c))^\ast)]&
\end{eqnarray}
The coefficient $X^R_\omega$ and $Y^R_\omega$ now depend on $R$ through the
potentials given by Eqs. (6.3)-(6.4), so that the perturbations propagate
differently in the 2
transverse directions. The explicit expressions $X^R_\omega(\sigma_c)$ and
$Y^R_\omega(\sigma_c)$ in terms of $\sigma_c$, obtained by combination of
Eqs. (6.11)-(6.12), (6.8)-(6.9), (6.3)-(6.4) and (5.11)-(5.12), are not very
enlightening. For convenience they are listed in the appendix. In the next
section we shall see, however, that it is possible to extract some simple
quantitative and qualitative physical consequences from these 2 functions.
\newpage
\section{Scattering formalism}
\setcounter{equation}{0}
In this section we will consider a scattering proces where a plane wave from
the asymptotic region $\sigma_c=-\infty$ is travelling along the string
towards the black hole. When the wave interacts with the gravitational field
of the black hole it will split into a reflected part returning to $\sigma_c=
-\infty$ and a transmitted part continuing towards the other asymptotic region
$\sigma_c=+\infty$ (see Fig.2.). The boundary conditions Eqs. (5.13) and (6.13)
were chosen such that up to second order we can identify the zeroth order
solution at $\sigma_c=-\infty$ with the solution in the asymptotic region
$\sigma_c=-\infty$:
\begin{equation}
\delta x_R^{(0-2)}(\sigma_c\rightarrow -\infty,\tau)=
\int d\omega\left( a^R_\omega e^{-i\omega(\tau-\sigma_c)}+
b^R_\omega e^{-i\omega (\tau+\sigma_c)}\right).
\end{equation}
In the other asymptotic region $\sigma_c=+\infty$ we find from Eqs.
(5.2)-(5.3), (5.14) and (6.14) up to second order in $\lambda$:
\begin{eqnarray}
\delta x_R^{(0-2)}(\sigma_c\rightarrow+\infty,\tau)=
\int d\omega[e^{-i\omega(\tau-\sigma_c)}(a^R_\omega(1+\bar{A}_\omega+
\bar{X}^R_\omega)\hspace*{-2mm}&+&\hspace*{-2mm}b^R_\omega(\bar{B}_\omega+
\bar{Y}^R_\omega)^\ast)\nonumber\\
+e^{-i\omega(\tau+\sigma_c)}(a^R_\omega(\bar{B}_\omega+\bar{Y}^R_\omega)+
b^R_\omega(1+\bar{A}_\omega\hspace*{-2mm}&+&\hspace*{-2mm}
\bar{X}^R_\omega)^\ast)],
\end{eqnarray}
where the bar indicates evaluation at $\sigma_c=+\infty$ in the functions
(5.11)-(5.12) and (6.11)-(6.12). Equation (7.2) can be written more compactly
as:
\begin{equation}
\delta x_R^{(0-2)}(\sigma_c\rightarrow+\infty,\tau)=\int d\omega[
c^R_\omega e^{-i\omega(\tau-\sigma_c)}+d^R_\omega e^{-i\omega(\tau+\sigma_c)}],
\end{equation}
where:
\begin{equation}
c^R_\omega=a^R_\omega(1+\bar{A}_\omega+\bar{X}^R_\omega)+b^R_\omega
(\bar{B}_\omega+\bar{Y}^R_\omega)^\ast,
\end{equation}
\begin{equation}
d^R_\omega=a^R_\omega(\bar{B}_\omega+\bar{Y}^R_\omega)+b^R_\omega(1+
\bar{A}_\omega+\bar{X}^R_\omega)^\ast.
\end{equation}
Eqs. (7.4)-(7.5) constitute the transformation giving the $(\sigma_c=+\infty)$-
amplitudes as a linear superposition of the $(\sigma_c=-\infty)$-amplitudes.
Energy conservation is then expressed as:
\begin{equation}
\mid a^R_\omega\mid^2-\mid b^R_\omega\mid^2=\mid c^R_\omega\mid^2-
\mid d^R_\omega\mid^2,
\end{equation}
which up to second order in $\lambda$ reads:
\begin{equation}
1=1+\mid\bar{A}_\omega\mid^2-\mid\bar{B}_\omega\mid^2+2Re(\bar{A}_\omega+
\bar{X}^R_\omega).
\end{equation}
{}From Eqs. (5.11)-(5.12) and (6.11) we find (taking $\omega>0$):
\begin{equation}
\bar{A}_\omega\equiv A_{\omega}(\sigma_c=+\infty)=0,
\end{equation}
\begin{equation}
\bar{B}_\omega\equiv B_{\omega}(\sigma_c=+\infty)=-i2\omega b K_o(2\omega b)
\frac{r_g}{b},
\end{equation}
\begin{equation}
Re(\bar{X}^R_\omega)\equiv Re(X^R_\omega(\sigma_c=+\infty))=2\omega^2 b^2
K^2_o(2\omega b)(\frac{r_g}{b})^2,
\end{equation}
so that Eq. (7.7) is indeed fulfilled. Here $K_o(x)$ is the Modified Bessel
function with the integral representation [8]:
\begin{equation}
K_o(x)=\int_0^\infty\frac{\cos xt}{\sqrt{1+t^2}}dt;\hspace*{1cm}x>0.
\end{equation}

Let us now consider a scattering proces where an ingoing wave is partly
reflected and partly transmitted, i.e. we look for solutions up to second
order in $\lambda$ in the form (see Fig.2.):
\begin{equation}
\delta x_R^{(0-2)} (\sigma_c,\tau)=\left\{ \begin{array}{cl}
a^R_\omega e^{-i\omega(\tau-\sigma_c)}+b^R_\omega e^{-i\omega(\tau+\sigma_c)}
&\mbox{for $\sigma_c\rightarrow-\infty$}\\
c^R_\omega e^{-i\omega(\tau-\sigma_c)}&\mbox{for $\sigma_c\rightarrow+\infty$}
\end{array}\right.
\end{equation}
$a^R_\omega$ is then the amplitude of the ingoing wave, $b^R_\omega$ is the
amplitude of the reflected wave and $c^R_\omega$ is the amplitude of the
transmitted wave. The reflection coefficient and transmission coefficient are
given by:
\begin{equation}
R^R_\omega=\mid\frac{b^R_\omega}{a^R_\omega}\mid^2,\hspace*{1cm}
T^R_\omega=1-R^R_\omega=\mid\frac{c^R_\omega}{a^R_\omega}\mid^2.
\end{equation}
Using Eqs. (7.4)-(7.5) with $d^R_\omega=0$ we get:
\begin{equation}
R^R_\omega=\mid\bar{B}^R_\omega\mid^2+2Re[\bar{B}_\omega(\bar{Y}^R_\omega)
^\ast]+{\cal O}(\lambda^4).
\end{equation}
The leading order of the reflection coefficient is therefore a second order
term independent of the polarization and given by (from Eq. (7.9)):
\begin{equation}
R^R_\omega=(2\omega b)^2 K^2_o(2\omega b)(\frac{r_g}{b})^2+{\cal O}(\lambda^3).
\end{equation}
The asymptotic behaviour of the Bessel function tells us [8]:
\begin{equation}
R^R_\omega\sim\left\{\begin{array}{cl}(\omega b)^2 e^{-2\omega b}&\mbox{for
$\omega b\gg 1$}\\(\omega b)^2\ln^2\omega b&\mbox{for $\omega b\ll 1$}
\end{array}\right.
\end{equation}
It follows that if the wavelength $\lambda_{-\infty}\equiv2\pi/\omega$ of
the free wave in the asymptotic region $\sigma_c=-\infty$
is considerably different from the
"impact parameter" $b$, then the reflection coefficient effectively vanishes
and the wave is almost completely transmitted.

To calculate the third order correction to $R^R_\omega$ we need the imaginary
part of $\bar{Y}^R_\omega$ in Eq. (7.14) since $\bar{B}_\omega$ is purely
imaginary. The evaluation of $\bar{Y}^R_\omega$ is quite involved (see the
appendix). After some tedious calculations we find:
\begin{equation}
Im(\bar{Y}^\perp_\omega)=-i\pi e^{-2\omega b}(\frac{7\omega b}{128}-
\frac{1}{16}-\frac{1}{32\omega b})(\frac{r_g}{b})^2+2\pi i\omega^2 b^2
K_o(2\omega b)(\frac{r_g}{b})^2,
\end{equation}
\begin{equation}
Im(\bar{Y}^\parallel_\omega)=Im(\bar{Y}^\perp_\omega)-\frac{i\pi}{32}
e^{-2\omega b}(\frac{3}{\omega b}+6+4\omega b)(\frac{r_g}{b})^2,
\end{equation}
from which we can write down explicit expressions for the third order
corrections to $R^R_\omega$ (using also Eq. (7.9)). It is more interesting,
however, to consider the difference in reflection coefficient in the 2
physical directions:
\begin{eqnarray}
\Delta R_\omega\hspace*{-2mm}&\equiv&\hspace*{-2mm}
R^\parallel_\omega-R^\perp_\omega=2Re[\bar{B}_\omega(\bar{Y}^\parallel_\omega-
\bar{Y}^\perp_\omega)^\ast]\nonumber\\
\hspace*{-2mm}&=&\hspace*{-2mm}\frac{\pi}{8}K_o(2\omega b)e^{-2\omega b}
(3+6\omega b+4\omega^2 b^2)(\frac{r_g}{b})^3.
\end{eqnarray}
This quantity is always positive so, not surprisingly, the reflection
coefficient for waves in the equatorial plane is larger than the reflection
coefficient for waves perpendicular to the equatorial plane, i.e. the waves
feel the black hole strongest when they are propagating along the string in
the equatorial plane. Finally we should mention that independently of the
relative size of the wavelength $2\pi/\omega$ and the "impact parameter"
$b$ (cf. the comment after
equation (7.16)) our reflection coefficients are always quite small when
evaluated as pure numbers. This is of course an artifact of our expansion
scheme which assumes $r_g/b\ll 1$, i.e. the gravitational field of the black
hole,that is
responsible for a non-vanishing reflection coefficient, is assumed to be
weak.

This concludes our investigations of the propagation of perturbations along a
string in the equatorial plane of a Schwarzschild black hole.
\vskip 12pt
{\bf Acknowledgements}\\
I would like to thank Valery Frolov for collaboration on the material presented
in Ref.1., part of which is reviewed here in sections 2-3, and for
helpfull discussions on other parts of the material presented here.
\newpage
\section{Appendix}
\setcounter{equation}{0}
In this appendix we give the explicit expressions for the functions
$X^R_\omega(\sigma_c)$ and $Y^R_\omega(\sigma_c)$ introduced in Eqs.
(6.11)-(6.12), and we present the calculations of $Re(\bar{X}^R_\omega)$
and $Re(\bar{Y}^R_\omega)$ used in Eqs. (7.10) and (7.17)-(7.18).
\vskip 6pt
\hspace*{-6mm}The explicit expressions for
$X^R_\omega(\sigma_c)$ and $Y^R_\omega(\sigma_c)$ are (section 6):
\begin{eqnarray}
X^R_\omega(\sigma_c)=\frac{r^2_g}{2i\omega}\int_{-\infty}^{\sigma_c}
\hspace*{-2mm}&[&\hspace*{-4mm}
\frac{5(b^2-2x^2)}{4(x^2+b^2)^3}-\frac{3x(3b^2-2x^2)}{2(x^2+b^2)^{7/2}}
\sinh^{-1}\frac{x}{b}\nonumber\\ \hspace*{-2mm}&-&\hspace*{-2mm}
\frac{b^2}{2(x^2+b^2)^3}\delta^R_\parallel-\frac{1}{8i\omega}\frac{x(2x^2-b^2)}
{(x^2+b^2)^4}\nonumber\\ \hspace*{-2mm}&-&\hspace*{-2mm}
\frac{e^{-2i\omega x}}{8i\omega}\frac{2x^2-b^2}{(x^2+b^2)^{5/2}}
\int_{-\infty}^x\frac{2t^2-b^2}{(t^2+b^2)^{5/2}}e^{2i\omega t}dt]dx
\end{eqnarray}
and:
\begin{eqnarray}
Y^R_\omega(\sigma_c)=-\frac{r^2_g}{2i\omega}\int_{-\infty}^{\sigma_c}
\hspace*{-2mm}&[&\hspace*{-4mm}
e^{2i\omega x}\frac{5(b^2-2x^2)}{4(x^2+b^2)^3}-e^{2i\omega x}\frac
{3x(3b^2-2x^2)}{2(x^2+b^2)^{7/2}}\sinh^{-1}\frac{x}{b}\nonumber\\
\hspace*{-2mm}&-&\hspace*{-2mm}\frac{b^2 e^{2i\omega x}}{2(x^2+b^2)^3}
\delta^R_\parallel-\frac{e^{2i\omega x}}{8i\omega}\frac{x(2x^2-b^2)}
{(x^2+b^2)^4}\nonumber\\
\hspace*{-2mm}&-&\hspace*{-2mm}\frac{1}{8i\omega}\frac{2x^2-b^2}
{(x^2+b^2)^{5/2}}\int_{-\infty}^x\frac{2t^2-b^2}{(t^2+b^2)^{5/2}}
e^{2i\omega t}dt]dx,
\end{eqnarray}
with the only polarization dependence through the $\delta^R_\parallel$-terms.
For the calculation of $\bar{X}^R_\omega$ and $\bar{Y}^R_\omega$ we will use
the following integrals [9]:
\begin{equation}
\int_0^\infty\frac{\cos px}{x^2+a^2}dx=\frac{\pi}{2b}e^{-pa},\nonumber
\end{equation}
\begin{equation}
\int_0^\infty\frac{\cos px}{\sqrt{x^2+a^2}}dx=K_o(pa),\nonumber
\end{equation}
\begin{equation}
\int_{-\infty}^{\infty}\frac{x\cos px}{(x^2+a^2)^{3/2}}\sinh^{-1}\frac{x}{a}
dx=\frac{\pi}{a}(e^{-pa}-paK_o(pa)),\nonumber
\end{equation}
as well as integrals obtained from them by differentiation with respect to
the constants $a$ and $p$. We will also use the equation for the Modified
Bessel function [8]:
\begin{equation}
x^2 K''_o(x)+xK'_o(x)-x^2 K_o(x)=0.\nonumber
\end{equation}
The real part of $\bar{X}^R_\omega$ used in Eq. (7.10) is now easily
calculated:
\begin{eqnarray}
Re(\bar{X}^R_\omega)\hspace*{-2mm}&=&\hspace*{-2mm}
\frac{r^2_g}{16\omega^2}\int_{-\infty}^{\infty}[\frac{2x^2-b^2}
{(x^2+b^2)^{5/2}}\cos 2\omega x\int_{-\infty}^x\frac{2t^2-b^2}{(t^2+b^2)^{5/2}}
\cos 2\omega t dt]dx\nonumber\\
\hspace*{-2mm}&=&\hspace*{-2mm}\frac{r^2_g}{32\omega^2}[\int_{-\infty}^{\infty}
\frac{2x^2-b^2}{(x^2+b^2)^{5/2}}\cos 2\omega x dx]^2\nonumber\\
\hspace*{-2mm}&=&\hspace*{-2mm}2\omega^2 b^2 K^2_o(2\omega b)(\frac{r_g}{b})^2.
\end{eqnarray}
The imaginary part of $\bar{Y}^R_\omega$ used in Eqs. (2.17)-(2.18) leads to:
\begin{eqnarray}
Im(\bar{Y}^R_\omega)\hspace*{-2mm}&=&\hspace*{-2mm}\frac{-r^2_g}{2i\omega}
\int_{-\infty}^{\infty}[\frac{5(b^2-2x^2)}{4(x^2+b^2)^3}-\frac{3x(3b^2-2x^2)}
{2(x^2+b^2)^{7/2}}\sinh^{-1}\frac{x}{b}]\cos 2\omega xdx\nonumber\\
\hspace*{-2mm}&+&\hspace*{-2mm}\frac{r^2_g b^2}{4i\omega}\delta^R_\parallel
\int_{-\infty}^{\infty}\frac{\cos 2\omega x}{(x^2+b^2)^3}dx+\frac{r^2_g}
{8i\omega^2}\int_{-\infty}^{\infty}\frac{x\sin 2\omega x}{(x^2+b^2)^4}
(2x^2-b^2)dx\nonumber\\
\hspace*{-2mm}&\equiv&\hspace*{-2mm}\alpha_\omega+
\beta_\omega\delta^R_\parallel+\gamma_\omega,
\end{eqnarray}
and from Eqs. (8.3)-(8.6):
\begin{equation}
\alpha_\omega=\frac{-\pi}{2i}[4\omega^2 b^2 K_o(2\omega b)+
\frac{e^{-2\omega b}}{8\omega b}(\frac{3}{4}+\frac{3}{2}\omega b-
\frac{15}{8}\omega^2 b^2)](\frac{r_g}{b})^2,
\end{equation}
\begin{equation}
\beta_\omega=\frac{\pi}{32i}e^{-2\omega b}(\frac{3}{\omega b}+6+
4\omega b)(\frac{r_g}{b})^2,
\end{equation}
\begin{equation}
\gamma_\omega=\frac{\pi}{64i}e^{-2\omega b}(\frac{1}{\omega b}+2-4\omega b)
(\frac{r_g}{b})^2,
\end{equation}
in agreement with Eqs. (7.17)-(7.18).
\newpage
\centerline{\bf References}\vskip 6pt\begin{enumerate}
\item A.L. Larsen and V.P. Frolov, Nordita Preprint 93/17 P
\item H.J. de Vega and N. S\'{a}nchez, Nucl. Phys. B309 (1988) 577
\item H.J. de Vega and N. S\'{a}nchez, Nucl. Phys. B309 (1988) 552
\item C.W. Misner, K.S. Thorne and J.A. Wheeler, Gravitation (Freeman,
      San Francisco, CA, 1973)
\item A.M. Polyakov, Phys. Lett. B103 (1981) 207
\item L.P. Eisenhart, Riemannian Geometry (Princeton University Press,
      fifth printing, 1964) Chapter IV
\item V.P. Frolov, V.D. Skarzhinsky, A.I. Zelnikov and O. Heinrich, Phys.
      Lett. B224 (1989) 255
\item M. Abramowitz and I.A. Stegun, Handbook of Mathematical Functions
      (Dover Publications Inc, New York, ninth printing) Chapter 9
\item I.S. Gradshteyn and I.M. Ryznik, Table of Integrals, Series and
      Products (Academic Press, Inc.)
\end{enumerate}
\newpage
\centerline{\bf Figure captions}
\vskip 6pt
\hspace*{-6mm}Fig.1. The potential $U_{\perp\perp}$
of Eq. (3.13) as a function of the
string parameter $\sigma_c$: $\sigma_c=0$ corresponds to the minimal
distance $r=m+\sqrt{m^2+b^2}$ between the stationary string and the black
hole, the 2 zeroes of $U_{\perp\perp}$ correspond to $r=m+\sqrt{m^2+3b^2/2}$
while the 2 maxima correspond to $\pm r=4m/3+\sqrt{16m^2/9+5b^2/2}$. The
potential $U_{\parallel\hspace*{.5mm}\parallel}$ has the same shape, but is
a little deeper near the black hole.
\vskip 12pt
\hspace*{-6mm}Fig.2. Schematic representation of the scattering processes
described in section 7: An ingoing wave from the asymptotic region
$\sigma_c =-\infty$ is partly
reflected and partly transmitted because of the interaction with the
gravitational field of the black hole. The circle represents the horizon of
the black hole in the equatorial plane.
\end{document}